\documentclass{ws-mplb}

\begin{document}

\markboth{Yao Yao, Yu-Zhong Zhang, Hunpyo Lee, et al.}{ORBITAL SELECTIVE PHASE TRANSITION}

\catchline{}{}{}{}{}
%
%%%%%%%%%%%%%%%%%%%%%%%%%%%%%%%%%%%%%%%%%%%%%%%%%%%%%%%%%%%%%%%%%%%%

\title{ORBITAL SELECTIVE PHASE TRANSITION}

\author{\footnotesize Yao Yao, Yu-Zhong Zhang$^*$}
\address{Shanghai Key Laboratory of Special Artificial Microstructure Materials and Technology, \\
School of Physics Science and Engineering, Tongji University, Shanghai 200092, P.R. China\\
$^*$yzzhang@tongji.edu.cn}

\author{\footnotesize Hunpyo Lee, Harald O. Jeschke, Roser Valent\'\i}
\address{Institut f\"ur Theoretische Physik, Goethe-Universit\"at Frankfurt, Max-von-Laue-Stra{\ss}e 1, 60438 Frankfurt am Main, Germany}

\author{\footnotesize Hai-Qing Lin}
\address{Beijing Computational Science Research Center, Beijing 100084, China}

\author{\footnotesize Chang-Qin Wu}
\address{Department of Physics and State Key Laboratory of Surface Physics, Fudan University, Shanghai 200433, China}

\maketitle

\begin{history}
\received{(Day Month Year)}
\revised{(Day Month Year)}
\end{history}

\begin{abstract}
  We review theoretical investigations on the origin of the orbital
  selective phase where localized and itinerant electrons coexist in the $d$ shell at intermediate strength of the on-site Coulomb interactions between electrons. In particular, the effect of spatial fluctuations on the phase diagram of the two-orbital Hubbard model with unequal bandwidths is discussed. And different band dispersions in different orbitals as well as  different
  magnetically ordered states in different orbitals which are responsible for orbital selective phase transitions are emphasized. This is due to the fact that these two
  mechanisms are independent of the Hund's rule coupling,
  and are completely distinct from other well-known mechanisms like orbitals
  of unequal bandwidths and orbitals with different
  degeneracies. Moreover, crystal field splitting is not required in
  these two recently proposed mechanisms.
\end{abstract}


%============================================
\section{introduction}
\label{sec:one}
%============================================

The one-band Hubbard model on various lattices has been extensively
investigated because of its complicated phase diagram which arises out
of the simple competition between kinetic energy and the on-site
Coulomb
interaction~\cite{Kyung2006PRL,Ohashi2006,Zhang2007,Meng2010,Sordi2012,Gull2013}. Though
topical exotic states like spin liquid and the high-$T_c$
superconducting state can be accounted for by this model, its
relevance to real materials is still limited due to the absence of
orbital degrees of freedom. This motivates a shift of attention
from the single-band Hubbard model to the multi-orbital one.

An orbital selective phase transition~\cite{Zhang2012}, which, in this
review, mainly refers to a  metal-to-insulator phase transition
that takes place at different critical values of the interaction
strength in different orbitals, is observed when orbital degrees of
freedom are taken into account in the Hubbard model. The phase
transition leads to an interesting intermediate state, so called
orbital selective phase, where localized spins and itinerant electrons
coexist in one system. Such a coexistence in a $d$ orbital was first
proposed by Anisomov {\it et al.}~\cite{Anisimov2002} and the scenario
was applied to explain exotic experimental findings in
Ca$_{2-x}$Sr$_x$RuO$_4$ at $0.2 \leq x \leq
0.5$~\cite{Nakatsuji2000}. Orbital selective phase transition
are believed to also occur in many other correlated systems like
iron-based
superconductors~\cite{Bascones2012,Medici2012,Yu2013,Yi2013} or
transition metal oxides~\cite{Shorikov2010,Huang2012} and is
generalized to involve other phase transitions like magnetic phase
transitions with orbital
dependence~\cite{Lee2010,Caron2012}. Furthermore, it is found to be
conceptually identical to the breakdown of the Kondo effect in
heavy-fermion metals~\cite{Pepin2007,Leo2008,Vojta2010}.

In this review, we will focus on the theoretical investigation of
possible sources for an orbital selective phase transition. The rest of the
paper is organized as follows. In Sec. \ref{sec:two},
materials which probably exhibit orbital selective phases are
discussed. In Sec. \ref{sec:three}, various mechanisms for an orbital
selective phase transition are presented. Conclusions are presented in
Sec. \ref{sec:four}.

%============================================
\section{Related Materials}
\label{sec:two}
%============================================

%============================================
\subsection{Orbital selective phase transition in Ca$_{2-x}$Sr$_x$RuO$_4$}
\label{sec:two-one}
%============================================

Ca$_{2-x}$Sr$_x$RuO$_4$ is a fascinating $4d$ multi-orbital system
that exhibits a rich and intricate phase diagram, ranging from a
chiral $p$-wave superconductor (Sr$_2$RuO$_4$) to a Mott insulator
(Ca$_2$RuO$_4$)~\cite{Nakatsuji2000,Maeno1994}. While it can be
inferred from susceptibility measurements that there is a local spin
$1/2$ present in the region of $0.2\leq x \leq 0.5$, the system
remains metallic and shows heavy fermion behavior as indicated from
transport studies~\cite{Nakatsuji2003}. In order to account for
this unusual phenomenon happening in the system without $f$ electrons,
an orbital selective phase transition was
proposed~\cite{Anisimov2002}: in Ca$_{2-x}$Sr$_x$RuO$_4$, there are
three degenerate $t_{2g}$ orbitals ($d_{xy}$, $d_{yz}$, and $d_{zx}$)
occupied by four $4d$ electrons. Due to the planar geometry, these bands
split into a wide, nearly two dimensional $d_{xy}$ band and two
narrow, nearly one-dimensional $d_{xz,yz}$ bands. By a combination of
local density approximation with dynamical mean field theory and
employing the noncrossing approximation as an impurity solver, successive
phase transitions were detected as a function of onsite interaction
strength $U$, from a metal where all the orbitals cross the Fermi
level to an orbital selective phase where a gap opens in the
$d_{xz,yz}$ orbitals while a strong peak in the density of states remains
at the Fermi level in the $d_{xy}$ orbital, and finally to an
insulator where all the orbitals become localized.

While the orbital selective phase transition offers a simple and clear
physical picture for the possible coexistence of localized and
itinerant electrons in $4d$ orbitals, the applicability of such a
scenario to Ca$_{2-x}$Sr$_x$RuO$_4$ is still questionable. Fang {\it
  et al.}~\cite{Fang2004} pointed out
by performing first-principles
bandstructure calculations that it is too naive to model
this system just by changing $U/t$ since in the real system
Ca$_{2-x}$Sr$_x$RuO$_4$,  rotation, tilting and flattening of the
octahedron play a crucial role for the understanding of the electronic
properties.  It was found that the strong RuO$_6$ rotation will
reduce the bandwidth of the $d_{xy}$ orbital significantly, but not
that of the $d_{yz}$, $d_{zx}$ orbitals.  This difference eventually
leads to a strong spin polarization and a pseudogap in the $d_{xy}$
orbital which dominantly contributes to the magnetization  while the
$d_{yz/zx}$ orbitals are quite broad and remain itinerant.  This
was in contrast
to the above suggested orbital-selective-phase-transition scenario.
In order to involve
the effect of octahedral distortions induced by Ca, Dai {\it et
  al.}~\cite{Dai2006} proposed a multi-orbital Hubbard model by
assuming the $d_{xy}$ band to be narrower than the $d_{yz/zx}$ band
and a negative crystal field splitting
$\Delta=\epsilon_{yz/zx}-\epsilon_{xy}$. Then, by applying slave boson
mean field calculations, the orbital selective phase transition is
recovered with $d_{xy}$ orbital being the first insulating band as $U$
is increased. However, Liebsch and Ishida~\cite{Liebsch2007}
introduced a new concept for the paramagnetic metal-to-insulator
transition in the layered perovskite Ca$_{2-x}$Sr$_x$RuO$_4$. They
noticed that the crystal field splitting between $d_{yz/zx}$ and
$d_{xy}$ states should be positive based on the observation of
increasing $d_{xy}$ orbital occupancy with increasing Ca concentration
from band structure calculations. Therefore, by using dynamical mean
field theory based on a finite temperature multiband exact
diagonalization, they showed that the combination of crystal field
splitting and on-site Coulomb interactions results in a complete
filling of the $d_{xy}$ orbital and a single Mott transition in the
half-filled $d_{yz/zx}$ bands, {\it i.e.}, no orbital selective phase
transition occurs in Ca$_{2-x}$Sr$_x$RuO$_4$. Recently, a local
density approximation plus dynamical mean-field theory with a
continuous time quantum Monte Carlo solver was applied to
Ca$_{2-x}$Sr$_x$RuO$_4$~\cite{Gorelov2010}, which allows for treating
realistically both the material-dependence and the local many-body
effects. In the metallic $x \leq 0.5$ phase, a progressive transfer of
electrons was found from the $d_{xy}$ orbital to the $d_{yz/zx}$
orbital, which is inconsistent with the positive crystal field
splitting picture, while down to $\sim 300$ K, no orbital selective
phase transition was detected.

Beside the debate of applicability of the orbital selective phase
transition scenario to Ca$_{2-x}$Sr$_x$RuO$_4$ at $0.2\leq x \leq
0.5$, controversies remain on the experimental side. For $x=0.2$, a novel
scenario of orbital selective phase transition was inferred from
angle-resolved photoemission experiments~\cite{Neupane2009}, while
other angle-resolved photoemission experiments show three metallic
bands and no orbital selective phase
transition~\cite{Shimoyamada2009}.

%============================================
\subsection{Orbital selective phase transition in iron-based superconductors}
\label{sec:two-two}
%============================================

Recently, the discovery of iron-based
superconductors~\cite{Kamihara2008,Johnston2010,Stewart2011} generated
extensive interest in the multi-orbital Hubbard model since all five
$3d$ orbitals of iron atoms cross the Fermi level and there is no way to
eliminate the orbital degrees of freedom and construct an effective
one-band model like what has been done in cuprates. Meanwhile,
electron-electron interaction is believed to play an essential role in
the superconductivity with high transition temperature as
electron-phonon coupling alone cannot explain the high critical value
of $T_c$ observed experimentally~\cite{Boeri2008}. Therefore,
a multi-orbital Hubbard model with intra- and inter-orbital Coulomb
interactions as well as Hund's rule coupling in addition to the
kinetic energy part is the minimal model for understanding the phase
diagrams of various iron-based
superconductors~\cite{Kuroki2008,Kaneshita2009}.

Superconductivity and  magnetism in iron-based
superconductors have been discussed both from  the
 strong coupling localized
limit~\cite{Si2008,Yildirim2008,Ma2009,Schmidt2010} -as usually done in
cuprates- and from  the  weak coupling itinerant
limit~\cite{Mazin2008,Zhang2009,Zhang2010,Zhang2011,Ding2013,Hirschfeld}. Both
scenarios are supported by some experimental works and theoretical
studies~\cite{Dai2012}. Recently, a compromise between these two
opposite points of view was proposed by assuming a coexistence of
localized spins and itinerant electrons in $3d$ orbitals, leading to a
double-exchange-like model~\cite{Yin2010,You2011}. While the model
offers a possible scenario for the origin of magnetism and
superconductivity in iron-based superconductors, the question why such
a model can be applied to these materials remains open. This is due to
the fact that the parent states of most high-$T_c$ iron-based
superconductors are distinct from the state of Ca$_{2-x}$Sr$_x$RuO$_4$
in the region of $0.2\leq x \leq 0.5$. For example, the former are
antiferromagnetic metals at low temperature without orbital degeneracy
while the latter is a paramagnetic metal with orbital
degeneracy. Furthermore, all the bandwidths of five $3d$ orbitals on
Fe atoms are almost equal while a remarkable difference can be seen in
the bandwidths of the three $t_{2g}$ orbitals on Ru atoms. These
differences indicate that the mechanism for orbital selective phase
transition discovered in Ca$_{2-x}$Sr$_x$RuO$_4$ cannot be directly
applied to iron-based superconductors. Thus,
if one follows this scenario,  a new mechanism for the
possible coexistence of local spins and itinerant electrons in $3d$
orbitals of Fe atoms is required in order to validate the application
of a double-exchange-like models to iron-based superconductors. In
fact, a temperature-induced crossover from a metallic state at low
temperatures to an orbital-selective phase at high temperatures was
experimentally observed in an angle-resolved photoemission
spectroscopy study on $A_x$Fe$_{2-y}$Se$_2$ ($A$=K, Rb)
superconductors~\cite{Yi2013}. And theoretically, various methods like
the Hartree-Fock approximation~\cite{Bascones2012},
slave-spin~\cite{Yu2013} or dynamical mean field
theory~\cite{Medici2012} have been employed to investigate the five
orbital Hubbard model. All these studies confirm that an orbital
selective phase transition should exist in a certain range of Hund's
rule coupling $J$ and on-site Coulomb interaction $U$. However, the
mechanism for the coexistence is still unknown~\cite{Zhang2012}.

%============================================
\subsection{Orbital selective phase transition in transition metal oxides}
\label{sec:two-three}
%============================================

Apart from the discussion of possible
orbital selective phase transitions in the above class of
topical materials, such a transition has been also proposed
in a few transition metal oxides.
 For example, in CoO~\cite{Huang2012}, high pressure
experiments revealed that there are two transitions in the resistivity.
  The first one happens around 60~GPa and
the second takes place around 130~GPa. However, room-temperature x-ray
emission spectroscopy studies on similar samples show that the spin
states of Co$^{2+}$ ions persist in the high spin state up to 140~GPa,
after which the crossover from a high spin state to a low spin state
happens. The second transition has a  close
relation to the spin state transition but  the first one is hard
to understand. Huang {\it et al.} performed density functional theory
calculations combined with dynamical mean field theory on CoO under
pressure and pointed out that the first transition  corresponds to
an orbital selective insulator-to-metal transition, where $t_{2g}$
orbitals of a Co $3d$ shell become metallic around 60~GPa while the
$e_g$ orbitals still remain insulating. A similar situation also occurs
in FeO~\cite{Shorikov2010}. Local density approximation plus dynamical
mean field theory calculations show that FeO at ambient pressure is an
insulator with a gap amplitude consistent with the experimental result and
at pressures higher than 60~GPa FeO is metallic but only for $t_{2g}$
while $e_g$ states remain insulating. This corresponds to an orbital
selective phase transition scenario. The result agrees with
high-pressure x-ray emission spectroscopy data~\cite{Badro1999}.

The concept of orbital selective phase transitions can
also be applied to the phase transitions in TiOCl under
pressure~\cite{Zhang2008}. By applying external hydrostatic pressure,
it is predicted from an ab initio molecular dynamics calculation that
the system undergoes consecutive phase transitions from a Mott
insulator to a metallic state through an intermediate phase where the
gap remains in the $d_{x^2-y^2}$ orbital due to the strong dimerization
between Ti atoms, while $d_{yz}$ and $d_{zx}$ orbitals start to cross
the Fermi level. However, further experiments are required to identify
the proposed intermediate phase in TiOCl.

%============================================
\subsection{Orbital selective phase transition in other systems}
\label{sec:two-four}
%============================================

Recently, the concept of orbital selective phase transition was
extended to a $^3$He bilayer system~\cite{Beach2011} which can
now be realized.  Experimentally,
  $^3$He films are grown on a
graphite/$^4$He bilayer substrate~\cite{Neumann2007}. The first layer
of $^3$He initially forms a monolayer fluid. With increasing $^3$He
content, a
second fluid layer forms. The resulting $^3$He fluid
bilayer comprises a nearly localized layer and an overlayer of
itinerant fermions. Below a characteristic temperature $T_0$, this
fluid bilayer has Fermi-liquid properties, with an enhanced
quasi-particle mass. The effective mass of the heavy-fermion state at
$T \ll T_0$ increases with increasing $^3$He density. Beyond a
critical $^3$He density $n_c$, the first layer is fully localized at
all temperatures. The system then comprises a solid $^3$He layer
forming an $S =\frac{1}{2}$ magnet on a triangular lattice, and a
fluid overlayer with relatively weak correlations and moderate
quasi-particle effective mass. This solidification of the first layer
can be interpreted as an orbital selective phase transition.

%============================================
\section{Theoretical models}
\label{sec:three}
%============================================

Though more experimental evidence is required for confirming the
existence of orbital selective phase transitions in real materials,
such a phenomenon
has  attracted intense interest from a theoretical point of view.
 In the
following, we will summarize related
 recent work on the theory of orbital
selective phase transitions. In order to include the orbital degrees
of freedom, a minimal two-orbital Hubbard model is used. If only the
paramagnetic state is taken into account, the dynamical mean field
theory~\cite{Georges1996} and its cluster extensions like dynamical
cluster approximation~\cite{Maier2005} are employed. And if
magnetically ordered states are considered, the Hartree-Fock
approximation is used and the results are cross-checked with the
two-sublattice dynamical mean field theory calculations.

%============================================
\subsection{Model and method}
\label{sec:three-0ne}
%============================================

The two-orbital Hubbard model~\cite{Zhang2012} is defined as
\begin{eqnarray}
&H&=-\sum_{\langle ij\rangle, \langle\langle ij\rangle\rangle, \gamma \sigma} t_{ij,\gamma} c^{\dagger}_{i\gamma\sigma}c_{j\gamma\sigma}
%-\sum_{\langle\langle ij\rangle\rangle \gamma\sigma} t_{\gamma}^{\prime} c^{\dagger}_{i\gamma\sigma}c_{j'\gamma\sigma}
+U\sum_{i\gamma}n_{i\gamma\uparrow}n_{i\gamma\downarrow} \nonumber \\
&+&\Big(U'-\frac J 2\Big)\sum_{i\gamma>\gamma'}n_{i\gamma} n_{i\gamma'}-2J\sum_{i\gamma>\gamma'}S_{i\gamma}\cdot S_{i\gamma'},\label{eq:hamiltonian}
\end{eqnarray}
where $t_{ij,\gamma}=t_{\gamma}$ ($t_{\gamma}^{\prime }$) is the
intra-orbital hopping integral between nearest-neighbor
(next-nearest-neighbor) sites denoted by $\langle ij\rangle$
($\langle\langle ij\rangle\rangle$) with orbital indices
$\gamma=\alpha,\beta$ in units of $t$. $U$, $U^{\prime }$ and $J$ are
the intra-band, inter-band Coulomb interaction and Hund's coupling,
respectively, which fulfill the rotational invariance condition
$U=U^{\prime }+2J$. The pair-hopping term ($-J\sum_{i,\gamma>\gamma'}
c^{\dagger}_{i\gamma\uparrow}c^{\dagger}_{i\gamma\downarrow}c_{i\gamma'\uparrow}c_{i\gamma'\downarrow}$)
is not explicitly written in the model (\ref{eq:hamiltonian}) since it
plays a minor role in the orbital selective phase transition as we will
show below. $c^{\dagger}_{i\gamma\sigma}$ ($c_{i\gamma\sigma}$)
creates (annihilates) an electron in orbital $\gamma$ of site $i$ with
spin $\sigma$. $n_{i\gamma\sigma}$ is the occupation operator, while
$n_{i\gamma}=n_{i\gamma\uparrow}+n_{i\gamma\downarrow}$, and
$S_{i\gamma}$ the spin operator.

The model was mainly solved by the dynamical mean field theory~\cite{Georges1996} which
is widely used in the research field of strongly correlated electron
systems. It maps a lattice many-body problem to an impurity one. This
mapping becomes exact if the local coordination becomes infinite, due
to the fact that the local and the impurity actions share the same
mathematical form in the limit of large coordination. Therefore, by
introducing the self consistent conditions, {\it i.e.}, that the local
Green's function and local self-energy are equivalent to the impurity
ones, respectively, a lattice many body problem can be solved after an
appropriate choice of impurity solver. In one of the studies on
the orbital selective phase transition done by some of the authors~\cite{Lee2010}, the Bethe lattice with
infinite coordination is used and the N\'eel antiferromagnetic state is
considered. In this case, the self consistent
conditions are given by
\begin{eqnarray}
 G_{0,A,\sigma}^{-1}=i\omega_n + \mu - t_{\gamma}^2 G_{B,\sigma} - t_{\gamma}^{\prime 2}G_{A,\sigma}, \\
 G_{0,B,\sigma}^{-1}=i\omega_n + \mu - t_{\gamma}^2 G_{A,\sigma} - t_{\gamma}^{\prime 2}G_{B,\sigma},
\end{eqnarray}
where $G_0^{-1}$ and $G$ are the Weiss fields and local Green's
function, respectively. Here A and B label two different sublattices
with opposite spins, $\omega_n$ is the Matsubara frequency, and $\mu$
is the chemical potential which controls the filling.

While dynamical mean field theory can deal with the dynamical
correlations in many body problems exactly, it completely neglects the
spatial fluctuations which are believed to be important in reduced
dimensions. In order to capture the effect of spatial correlations,
cluster extensions of dynamical mean field theory were proposed, where
the single site embedded into the medium is replaced by a
cluster and spatial fluctuations are therefore partially considered.
 The larger the cluster is, the better the description
of spatial correlations is. Here studies of the
two orbital Hubbard model on the square lattice by the dynamical cluster approximation~\cite{Maier2005}, one of the cluster extensions of dynamical mean field theory, will be detailedly reviewed. The self consistent
condition reads as follows:
\begin{eqnarray}
G_C(\bf{K},i\omega_n) &=& \frac{N_c}{N_{t}} \sum_{\tilde{\bf{k}}}
\frac{1}{i\omega_n + \mu - \epsilon_{\tilde{\bf{k}}+\bf{K}} -\Sigma(\bf{K},i\omega_n) } \\
&=&
\frac{1}{G_0^{-1}(\bf{K},i\omega_n) -\Sigma(\bf{K},i\omega_n) },
\nonumber
\end{eqnarray}
where $G_C$ is the cluster Green's function for both the lattice and
the impurity models, and $\bf{K}$ labels a cluster wave number,
$\tilde{\bf{k}}$ denotes the lattice wave numbers in the Wigner-Seitz
cell surrounding $\bf{K}$, and $N_t$ is the total number of lattice
sites. $G_0(\bf{K},i\omega_n)$ is the Weiss field. If the cluster size of
$N_c$ is chosen to be $1$, $\bf{K}$ to be $(0,0)$, and
$\tilde{\bf{k}}$ runs over the whole Brillouin zone, the
self-consistency condition of dynamical mean field theory is
recovered. This $N_c=1$ dynamical cluster approximation can be used to verify the
validity of the Hartree-Fock approximation calculations. In the following, we mainly focus on the studies where four-site cluster
which preserves the $C_4$ symmetry of the square lattice is used.

In the recent studies done by some of the authors~\cite{Zhang2012,Lee2010,Lee2010PRL,Lee2011Ann,Lee2011}, a numerically exact method, the
weak-coupling continuous time quantum Monte Carlo method, is used as an
impurity solver~\cite{Gull2011}, which expands the Feynman diagrams in
terms of the interaction $U$. Compared to other numerically exact solvers,
this solver become computationally more feasible than the
strong-coupling continuous time quantum Monte Carlo~\cite{Gull2011}
when the local
Hilbert space of the cluster becomes large, and it can work at low
temperatures without the Suzuki-Trotter decomposition error when compared to
the Hirsch-Fye quantum Monte Carlo~\cite{Georges1996}.
 Moreover, the solver treats the
lower and upper Hubbard band as well as the quasiparticle peak at the same
level of
precision whereas, for instance,
 the numerical renormalization group method~\cite{Bulla2008} works only
perfectly close to the Fermi level.

\begin{figure}[tb]
\includegraphics[width=0.96\textwidth]{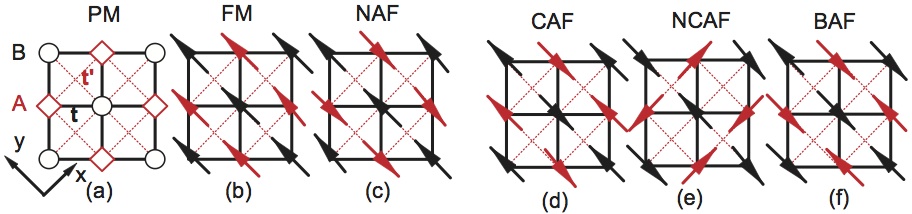}
\caption{(Color online) Cartoons for the different magnetically
  ordered states. (a) Paramagnetic (PM)
  state. The choices of sublattice and coordinate system are
  shown. (b) Ferromagnetic {FM}, (c) N\'eel (NAF), (d) collinear
  (CAF), (e) noncollinear {NCAF}, (f) bi-collinear (BAF)
  antiferromagnetic states. (From Ref. 7.)} \label{fig:pattern}
\end{figure}

Finally, in order to access various magnetic states in a uniform
formulation of the two dimensional system with hoppings up to
 next-nearest-neighbor sites, the Hartree-Fock approximation was applied by some of the authors to
the two-orbital Hubbard model on the frustrated square
lattice~\cite{Zhang2012}. The corresponding cartoons for different magnetic patterns
are shown in Fig.~\ref{fig:pattern}~(a)-(f).

%============================================
\subsection{Mechanism for orbital selective phase transition}
\label{sec:three-two}
%============================================

After the first proposal of an orbital selective phase transition
scenario in Ca$_{2-x}$Sr$_x$RuO$_4$, much effort has been devoted to
understand the origin of this transition and its
theoretical description. Various theoretical
approaches have been used.
For example, slave spin mean field theory~\cite{Medici2005}, the Gutzwiller variational
approach~\cite{Ferrero2005}, the determinant quantum Monte Carlo method~\cite{Bouadim2009},
in addition to
the dynamical mean field theory with different impurity solvers like
exact diagonalization, Hirsch-Fye quantum Monte Carlo, continuous
quantum Monte Carlo, etc. In the following, we will review five
possible mechanisms for the orbital selective phase transition
currently existing in the literature.

%============================================
\subsubsection{Different orbitals with different bandwidths at half filling}
\label{sec:three-two-one}
%============================================

It was originally proposed by Anisomov {\it et al.} that orbitals with
unequal bandwidths are the mechanism for the orbital selective phase
transition~\cite{Anisimov2002}. However, Liebsch~\cite{Liebsch2003}
 questioned the
existence of the transition by performing a dynamical mean field theory
calculation combined with a finite-temperature Hirsch-Fye quantum Monte
Carlo method for the two-orbital model without spin-flip term using
$J=U/4$, $U'=U/2$, and semi-elliptical densities of states with a
bandwidth ratio of $W_2/W_1=0.5$.  Additional
studies using the dynamical mean field theory with iterative
perturbation theory seemed to confirm Liebsch's
 conclusion of a single Mott
transition of both bands at the same critical
$U$-value~\cite{Liebsch2004}. In contrast, Koga {\it et
  al.}~\cite{Koga2004} found an orbital selective phase transition in
the two orbital model including additional pair hopping terms.
These authors  considered
exact diagonalization as the impurity solver in their dynamical mean
field theory calculations. Consequently, the orbital selective phase
transition scenario was attributed to spin-flip and pair-hopping
processes.

In a subsequent study, Ferrero {\it et al.}~\cite{Ferrero2005} applied the
Gutzwiller variational approach to the two orbital model with pair
hopping terms at temperature $T=0$ and confirmed the existence of an
orbital selective phase transition, provided that the ratio $W_2/W_1$
of the two bandwidths is sufficiently small. Similar results were
obtained by de' Medici {\it et al.}~\cite{Medici2005}, who used
slave-spin mean field theory (which is closely related to the
Gutzwiller method).  Arita and Held~\cite{Arita2005} considered the
dynamical mean field theory combined with the projective quantum Monte
Carlo method to investigate the same model at $T=0$ and demonstrated
an orbital selective phase transition for $J=U/4$ and $U=2.6$ (in
units of half the narrow-band width). Remarkably, a different
conclusion was drawn by Knecht {\it et al.}~\cite{Knecht2005}
who investigated
 the two orbital Hubbard model without spin flip term. By performing
 dynamical mean field theory calculations with a refined
Hirsch-Fye quantum Monte Carlo method, these authors
found two consecutive orbital
selective phase transitions, which separate a metallic
phase from an insulating state with an intermediate orbital selective
phase where  localized and itinerant electrons coexist. Later on,
a similar conclusion was found by Liebsch~\cite{Liebsch2005} who performed
 dynamical mean field theory calculations with a finite temperature
exact diagonalization impurity solver.
  Finally, in order to study the nonlocal correlation effects on
the orbital selective phase transition, Bouadim {\it et
  al.}~\cite{Bouadim2009} applied the determinant quantum Monte Carlo
method to a ferromagnetic Kondo lattice model which can be viewed as
a two-orbital Hubbard model without spin-flip term
where the electrons in one orbital are fully localized.
An orbital selective phase was confirmed to exist
even with the inclusion of nonlocal correlations. However, the
consequences of involving the nonlocal correlations directly on the two
orbital Hubbard model remained unknown.

\begin{figure}
\begin{center}
\includegraphics[width=0.80\textwidth]{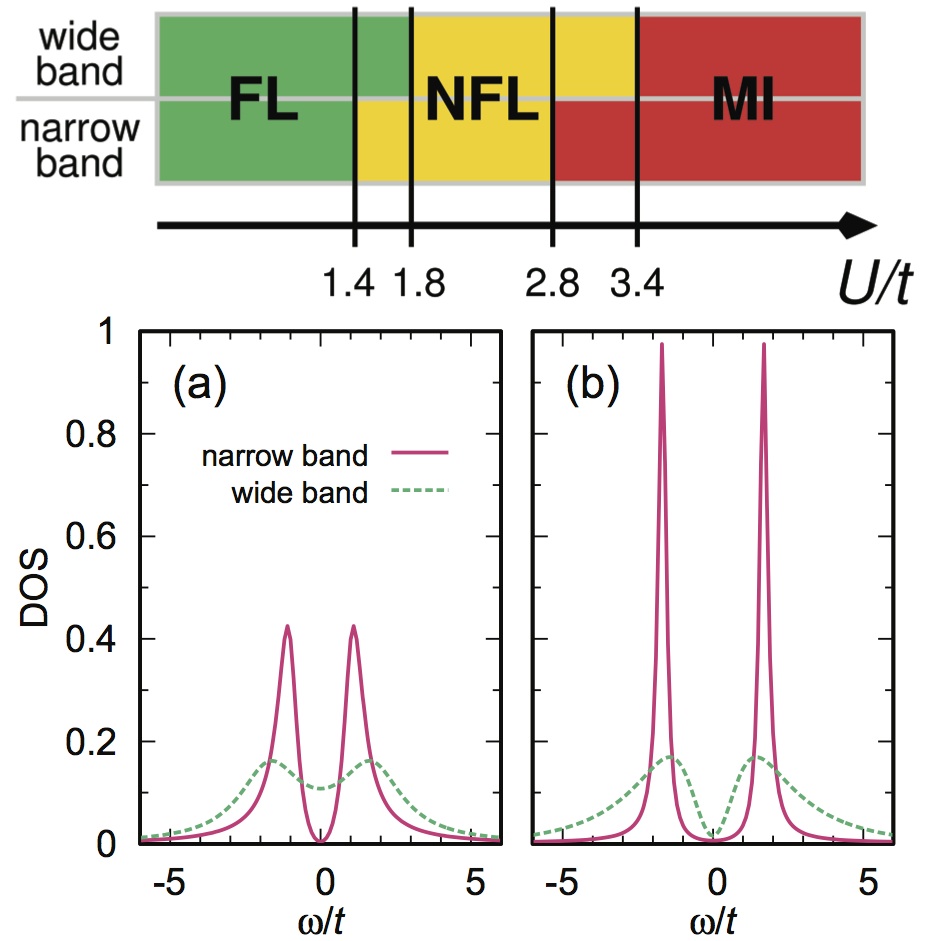}
\end{center}
\caption {(Color online) Upper panel is the phase diagram with five
  phases. Abbreviation FL stands for Fermi liquid, NFL for non-Fermi
  liquid, and MI for Mott insulator. The right panel is the density of
  states for (a) $U/t=2.8$ and (b) $U/t=3.4$. Here $J/U=0.25$,
  $U'/U=0.5$, $T/t=0.1$ and cluster size $N_c=4$, and the
  Pad{\'e} approximation method is employed for the analytic continuation. (From
  Ref. 65.)} \label{fig:DCA_diffbandwidth_PD}
\end{figure}

In order to explore the effect of nonlocal correlations
in the two orbital Hubbard model (without spin flip term),
Lee. {\it et al.} considered the dynamical
cluster approximation with cluster sizes up to $N_c=4$ and the weak coupling
continuous time quantum Monte Carlo as an impurity solver~\cite{Lee2010PRL,Lee2011Ann}. The upper panel of
Fig.~\ref{fig:DCA_diffbandwidth_PD} shows the resulting phase diagram
which shows a more complex structure
than what we learnt from previous
investigations. Here the ratio of $J/U=0.25$, $U'/U=0.5$ and
the temperature of $T/t=0.1$ is fixed. Though only the nonlocal spatial
correlations within a 4-site cluster as well as the dynamical
correlations were included, previous works discussed
that
the main correlations can already be captured with such a minimal
choice of cluster size~\cite{Kyung2006,Sakai2012}.

The phase diagram shows
 five distinctive phases with
two orbital selective phases. One exists from $U/t=2.8$ to $U/t=3.4$, which is characterized by a coexistence of localized electrons in the
narrow band and itinerant electrons in the
wide band. Interestingly,
 the metallic state in the wide band is of non-Fermi liquid behavior.
 It will be shown later how the nature of metallic states are determined.
 The phase transition from this orbital selective phase to the
 insulating state can be clearly seen in the variation
 of the density of state as the interaction $U$ increases.
 In the lower panel of Fig.~\ref{fig:DCA_diffbandwidth_PD}, it is showed that, while a gap is present in the narrow band at $U/t=2.8$, finite density of states remains at the Fermi level in the wide band. Please note,
that the quasiparticle peak vanishes at the Fermi level, which hints
to a non-Fermi liquid behavior in wide band. At $U/t=3.4$, a gap starts to open also in the wide band. The second orbital selective phase located
between $U/t=1.4$ and $U/t=1.8$ is of a novel
type.  It is characterized by a coexistence of two different metallic states,
 {\it i.e.}, Fermi liquid in the
wide band and  non-Fermi liquid in the narrow band. Besides
 these two orbital selective phases, there are three phases called metallic states with Fermi liquid behavior in both orbitals, metallic states with non-Fermi liquid behavior in both orbitals, and insulators in both orbitals.

\begin{figure}
\includegraphics[width=0.96\textwidth]{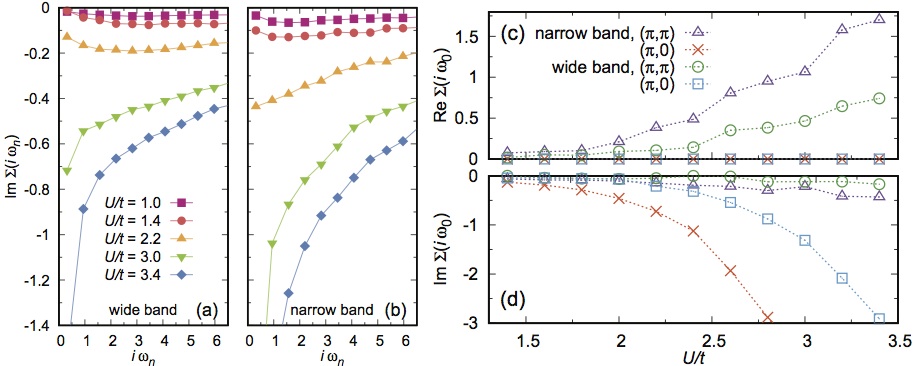}
\caption {\label{Nc4_selfenergy} (Color online) The imaginary part of the on-site
  self-energy for $U/t=1.0$,1.4,2.2,3.0,3.4 at $J/U=0.25$, $U'/U=0.5$, and $T/t=0.1$ for $N_c=4$
  (a) in the narrow band and (b) in the wide band. The real (c) and
  imaginary (d) part of self-energy at the lowest Matsubara frequency
  $\omega_0$ for $K$=($\pi$,$\pi$) and ($\pi$,0) sectors as a function
  of $U/t$. (From Ref. 65.)} \label{fig:DCA_diffbandwidth_self}
\end{figure}

The nature of the metallic states in general
can be obtained from the analysis of self-energy. The imaginary part of the on-site self-energy $\rm{Im}\,\Sigma (i\omega_n) = \sum_K \rm{Im}\,\Sigma (K,i\omega_n)$ provides information about the possible Fermi-liquid/non-Fermi-liquid behavior of the system as well as the nature of the gap opening. In Figs.~\ref{fig:DCA_diffbandwidth_self}(a) and (b) $\rm{Im} \,\Sigma (i\omega_n)$ is presented for the narrow and wide bands, respectively,
 at sufficiently low temperature of $T/t=0.1$.
 According to Fermi-liquid theory, $\rm{Im}\,\Sigma (\omega)$ at $T=0$
 extrapolates to $0$ at $\omega \rightarrow 0$, indicating that quasiparticles
 have an  infinite life time at the Fermi level. In the weak-coupling regions below $U/t=1.4$, such as at $U/t=1.0$, this Fermi-liquid behavior is seen in both bands. Between $U/t=1.4$ and $1.8$, like $U/t=1.4$, Fermi-liquid behavior is still present in the wide band, while non-Fermi-liquid behavior is observed in the narrow band. The quasiparticles at the Fermi level begin to have finite life time due to the finite value of the imaginary part of the self-energy at $i\omega_n \rightarrow 0$ driven by the nonlocal correlations, while those in the wide band still have infinite life time. As the interaction is increased, for example at $U/t=2.2$, non-Fermi-liquid behavior is observed in both bands. At
 $U/t=3.0$, $\rm{Im}\,\Sigma (i\omega_n)$ in the narrow band diverges, which indicates the opening of a Mott gap due to the divergence of scattering rate at the Fermi level, while the metallic state (non-Fermi-liquid) is still present in the wide band. These results evidence an orbital selective phases transition. In the strong-coupling region, for example at $U/t=3.4$, the Mott insulating state is observed in both bands, as suggested by the divergent behaviors of $\rm{Im}\,\Sigma (i\omega_n)$ in both bands.

Next, the nature of the gap opening was analyzed by scrutinizing the real and imaginary part of self-energy at the
smallest Matsubara frequency in different momentum
 sector as a function of $U/t$. Already at the level
of the one-band Hubbard model on the square lattice solved
with the  cluster dynamical mean field theory  with 4-site
cluster sizes and the
 strong-coupling continuous time quantum Monte Carlo
method~\cite{Park2008}, different nature of insulating
behavior was observed in different momentum
sectors  $K$=$(0,0)$/$(\pi,\pi)$ and $(\pi,0)$/$(0,\pi)$.
 In Figs.~\ref{fig:DCA_diffbandwidth_self} (c) and (d), respectively, the real and imaginary parts of the self-energy are presented at the lowest Matsubara frequency $\omega_0$, $\rm{Re}\,\Sigma(i\omega_0)$ and $\rm{Im}\,\Sigma(i\omega_0)$, for $K$=($\pi$,$\pi$) and ($\pi$,0) in both bands.  While $\rm{Re}\,\Sigma(i\omega_0)$ gives information about the energy shift of the spectral function, $\rm{Im}\,\Sigma(i\omega_0)$ introduces the scattering rate. As the interaction is increased, $\rm{Re}\,\Sigma_{ (\pi , \pi)}=-\rm{Re}\,\Sigma _{(0,0)}$ increases while $\rm{Im}\,\Sigma_{(\pi , \pi)}=\rm{Im}\,\Sigma_{ (0 , 0)}$ remains small in both bands. These results suggest a band insulator in the momentum sectors $K$=$(0,0)$ and $(\pi,\pi)$ where the formation of the gap is due to the separation of the poles. On the other hand, as the interaction increases $\rm{Im}\,\Sigma_{ (\pi , 0)}=\rm{Im}\,\Sigma_{ (0 , \pi)}$ displays a divergent behavior and $\rm{Re}\,\Sigma_{ (\pi ,  0)}=-\rm{Re}\,\Sigma_ {(0 , \pi)}$ in both bands is zero due to the particle-hole symmetry. Therefore, in the strong coupling region, the gap in the $K$=$(\pi ,0)$ and $(0, \pi)$ sectors is only induced by the divergence of $\rm{Im}\,\Sigma(i\omega_0)$ which is a signature for Mott physics. These results are similar to the single-band Hubbard model results~\cite{Park2008} but, while a first-order transition occurs in the single-band Hubbard model,
an orbital selective phase transition happens in the two-band Hubbard model.

\begin{table}[pt]
\tbl{Nearest ($t_{\gamma}$) and next-nearest neighbor ($t_{\gamma}'$) intra-orbital hoppings derived from density functional theory calculations on the iron-based superconductor LaOFeP through construction of the Wannier orbitals. }
{\begin{tabular}{@{}ccccccc@{}} \toprule
&$\gamma=$&   $d_{xy}$   &   $d_{yz}   $ &   $d_{3z^2-r^2}   $ &   $d_{zx}$   &   $d_{x^2-y^2}   $ \\ \colrule
$t_{\gamma} (meV)$ & & -342 & 231 & 130 & 231 & -257 \\
$t_{\gamma}'$  (meV) & & 106 & 268\tabmark{a} &  22 & 268\tabmark{a} &  156 \\ \colrule
$|t_{\gamma}'/t_{\gamma}|$ & & 0.31 & 1.16 & 0.17 & 1.16 & 0.61 \\ \botrule
\end{tabular} }\label{tab:hoppings}
\begin{tabfootnote}
\tabmark{a} The next-nearest neighbor hoppings $t_{\gamma}'$ with $\gamma=d_{yz}$ and $d_{zx}$ are obtained by an average over the corresponding hoppings in $[1,0,0]$ and $[0,1,0]$ directions~\cite{Miyake2010}.
\end{tabfootnote}
\end{table}

%============================================
\subsubsection{Different orbitals with different band dispersions}
\label{sec:three-two-two}
%============================================

While the mechanism of unequal bandwidth is thoroughly studied in the literature, the effect of details of the band dispersions of different orbitals
has been  less investigated. Its importance was
 only recognized after the discovery of iron-based superconductors~\cite{Lee2010,Zhang2011}. As shown in Table 1, the strength of the frustration, defined by the ratio of nearest and next-nearest neighbor intra-orbital hoppings $t_{\gamma}'/t_{\gamma}$~\cite{Miyake2010}, is very distinctive in different orbitals, {\it i.e.}, some of the orbitals like $d_{yz}$ and $d_{zx}$ exhibit strong frustration while others weaker, indicating that the dispersion relations of different orbitals are quite different. As we mentioned above, the iron-based superconductors are assumed to be possible
 candidates showing orbital selective phase
 where localized and itinerant electrons coexist.
However, a mechanism based on orbitals with unequal bandwidths
 is not applicable here since the bandwidths of five $3d$ orbitals are almost the same. In the present case,
 orbitals with different band dispersions could be
 a possible origin for the orbital selective phase transition.
 As most of the iron-based superconductors are magnetically
 ordered at low temperature, this issue was studied by allowing for magnetic solutions.

\begin{figure}
\includegraphics[width=0.96\textwidth]{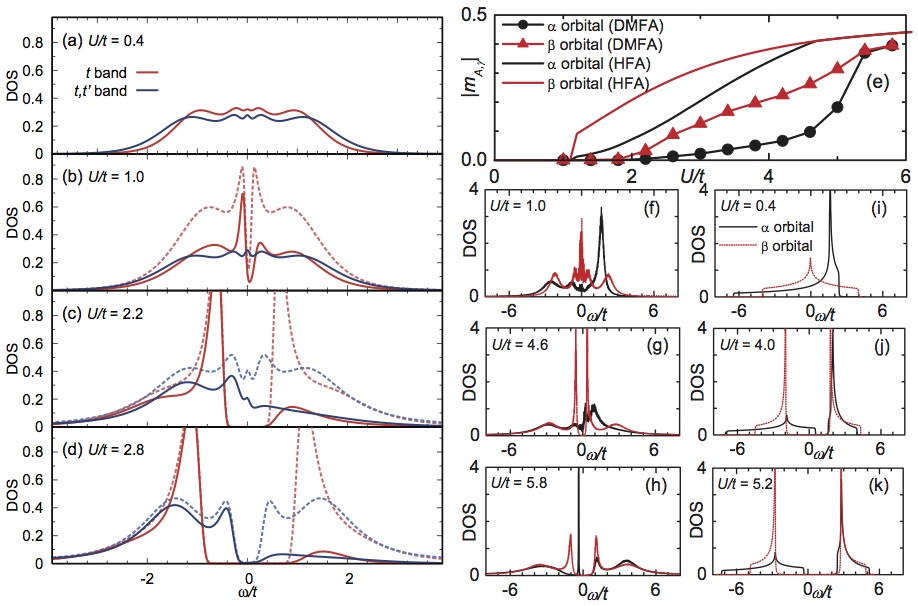}
\caption{(Color online) Left panel: Spin-up DOS on sublattice A (solid lines) as a function of
  frequency analyzed at $J/U=0.25$, $U'/U=0.5$, and $T/t = 1/32$ for (a) $U/t=0.4$, (b) $U/t=1.0$,
  (c) $U/t=2.2$ and (d) $U/t=2.8$. Also shown (dotted lines) is the
  total spin-up DOS when antiferromagnetic order occurs. (From Ref. 16.) Right panel: Comparison of the results from Hartree-Fock
  approximation (HFA) and dynamical mean-field approximation (DMFA) at
  $t_\alpha=1$, $t_\alpha'=0.6$, $t_\beta=1$, $t_\beta'=0$, $J/U=0.25$, and $U'/U=0.5$. e) Magnetization as a function of $U/t$. (f)-(h)
  ((i)-(k)) show the density of states in different phases from DMFA
  (HFA). (From Ref. 7.)}\label{fig:DMFT_diffbandstruct_DOSComp}
\end{figure}

In order to capture the underlying physics,
Lee {\it et al.}~\cite{Lee2010} applied a two sublattice dynamical mean field theory with the weak-coupling continuous time Quantum Monte Carlo method to the two-orbital Hubbard model with one orbital frustrated and the other unfrustrated which should mimic the effect of coupling between weakly and strongly frustrated bands in Fe-based superconductors. The N\'eel ordered spin arrangement is used. The interaction parameters are fixed at $U'=\frac{U}{2}$ and $J=\frac{U}{4}$ and the spin-flip term is ignored in model (\ref{eq:hamiltonian}). The calculations were firstly performed on the Bethe lattice. Here $t_1=1$ and $t_1'=0$ is set for the unfrustrated band and $t_2=1$ and $t_2'=0.65$ for the frustrated one. The bandwidths for unfrustrated and frustrated bands were
 $W_1=4.0$ and $W_2=4.77$, respectively.

Fig.~\ref{fig:DMFT_diffbandstruct_DOSComp} shows the spin-up density of state on the A site for four representative values of $U/t$ at a fixed temperature of $T/t=1/32$. At $U/t=0.4$, both bands are of the paramagnetic metallic state (see Fig.~\ref{fig:DMFT_diffbandstruct_DOSComp}~(a)). When the interaction $U/t$ is increased to $1.0$ (see Fig.~\ref{fig:DMFT_diffbandstruct_DOSComp}~(b)), the frustrated band ($t,t'$ band) remains in a paramagnetic state while a pseudogap is present in the unfrustrated band ($t$ band). As we will see below that the opening of a pseudogap can be attributed to a small moment appearing in the unfrustrated band.  As the interaction is further increased, an orbital selective phase transition occurs, and at $U/t=2.2$ (see Fig.~\ref{fig:DMFT_diffbandstruct_DOSComp}~(c)), a metal in the frustrated band coexists with an insulator in the unfrustrated band. Finally, in the strong-coupling region, for example at $U/t=2.8$ (see Fig.~\ref{fig:DMFT_diffbandstruct_DOSComp}~(d)), both bands are in insulating states.

\begin{figure}
\includegraphics[width=0.96\textwidth]{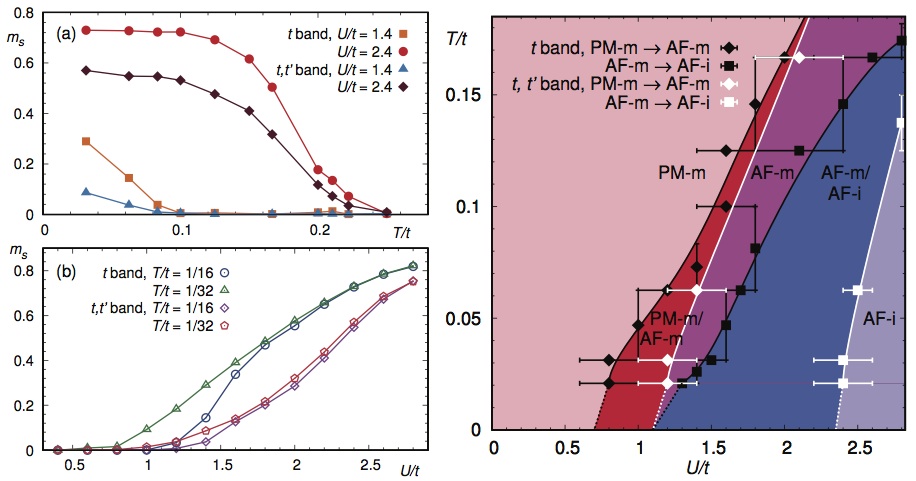}
\caption{(Color online) Left panel: (a) Staggered magnetization $m_s$ for the unfrustrated band
  ($t$ band) and the frustrated band ($t,t'$ band) (a) as a function
  of $T/t$ for $U/t=1.4$ and $2.4$ and (b) as a function of $U/t$ for
  $T/t=1/16$ and $1/32$.  A continuous transition with a smooth
  increase of $m_s$ is observed as a function of $U/t$. Right panel: Magnetic phase diagram for the two-band Hubbard model where
  an unfrustrated band ($t$ band, $t=1$,$t'=0$) and a frustrated band
  ($(t,t')$ band, $t=1$, $t'=0.65$) coexist. The phase boundaries'
  error bars are also shown. Abbreviations PM and AF denote paramagnetic and antiferromagnetic, respectively, while M (I) is for metal (insulator). Here $J/U=0.25$ and $U'/U=0.5$. (From Ref. 16.)}\label{fig:DMFT_diffbandstruct_MP}
\end{figure}

In order to understand the origin of the pseudogap behavior,
the staggered magnetization $m_s$ was studied as a function of temperature $T/t$ and interaction $U/t$. In the left panel of Fig.~\ref{fig:DMFT_diffbandstruct_MP}~(b) the staggered magnetization is shown as a function of interaction strength $U/t$ for two temperature values. A smooth increase of the magnetization is found with increasing $U/t$ for both bands and for both temperatures, indicating that the paramagnetic to antiferromagnetic phase transitions are of continuous order. Combining the results from the calculation of density of states, it suggests the existence of an antiferromagnetic metal where the small staggered magnetization is not sufficient for opening a full gap. In left panel of Fig.~\ref{fig:DMFT_diffbandstruct_MP}~(a) the behavior of the staggered magnetization $m_s$ is shown as a function of temperature $T/t$ for two different interaction strengths $U/t$. At $U/t=1.4$ ($U/t=2.4$), the staggered magnetization $m_s>0$ for both frustrated and unfrustrated bands is detected as temperature decreases below the N\'eel temperature around $T_N/t\simeq0.1$ ($T_N/t\simeq0.22$) where the system undergoes a paramagnetic to antiferromagnetic phase transition. The staggered magnetization increases more rapidly in the unfrustrated band than in the frustrated one.

In the right panel of Fig.~\ref{fig:DMFT_diffbandstruct_MP} the phase diagram $T/t$ versus $U/t$ is plotted for the Hamiltonian~(\ref{eq:hamiltonian}) without spin-flip term. The paramagnetic metal, antiferromagnetic metal and antiferromagnetic insulator are present in both bands, but the critical values $U_c/t$ of the unfrustrated band are smaller than those of the frustrated one. This leads to a complex phase diagram with five phases appearing. For example, at $T/t=0.05$, consecutive phase transitions are obtained as the interaction $U/t$ increases, first from paramagnetic metals in both bands to a new orbital selective phase where a paramagnetic metal coexists with an antiferromagnetic metal, then to a state with both bands of antiferromagnetic metallic states. Further increasing $U/t$, an orbital selective phase occurs with an antiferromagnetic metal in frustrated orbital ($t$,$t'$ band) while an antiferromagnetic insulator in unfrustrated orbital ($t$ band). Finally, both bands becomes antiferromagnetic insulators through the orbital selective phase transition.

\begin{figure}[tb]
\includegraphics[width=0.96\textwidth]{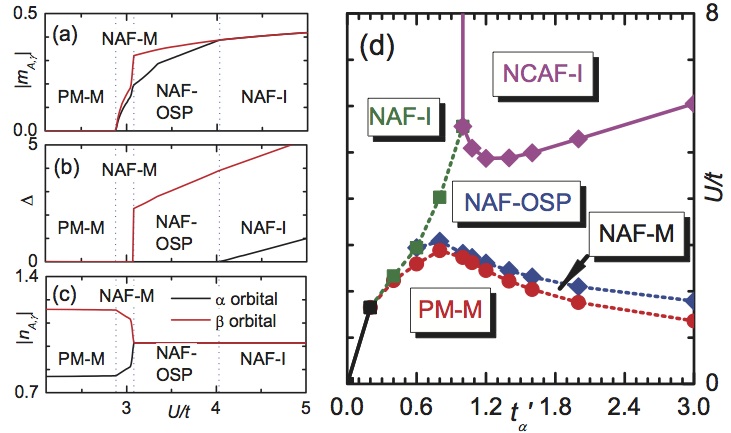}
  \begin{center}
\caption{(Color online) Variation of the orbital dependent magnetization (a), gap amplitude (b), and occupation number (c) as a function of $U/t$ at $t_\alpha=1$, $t_\alpha'=0.8$ and $t_\beta=1$, $t_\beta'=0$.
(d) Phase diagram in $U/t$-$t_\alpha'$ at $t_\alpha=1$, $t_\beta=1$, $t_\beta'=0$. Here $J/U=0.25$, $U'/U=0.5$ and filling is $1/2$. Regions of different phases are indicated by the abbreviations, for instance, paramagnetic by PM, N\'{e}el antiferromagnetic by NAF, noncollinear antiferromagnetic by NCAF, while orbital selective phase by OSP. M (I) denotes metal (insulator). Solid and dotted lines in the phase diagram represent first and second order phase transitions, respectively. (From Ref. 7.)} \label{fig:HF_diffbandstruct_PD}
  \end{center}
\end{figure}

While a phase diagram including orbital selective phase transitions was obtained within the dynamical mean field theory calculations, some fundamental issues concerning the origin of the detected orbital selective phase transition could not be explicitly addressed due to the multitude of possibilities coexisting in
the above study, such as different bandwidths, variation of filling factor as a function of Coulomb interaction due to the fixed chemical potential, different ratios of next-nearest and nearest neighbor hoppings in different orbitals, as well as dynamical fluctuations. Also, various simplifications employed in the
above study had to be justified.  For instance, the constraint of N\'eel antiferromagnetic order, the artificially imposed particle-hole symmetry even when next-nearest neighbor hoppings are involved in infinite dimensions, and the neglect of the spin-flip terms in the Hund's rule coupling. Furthermore, other parameter regimes like doping and crystal field splittings could not be clarified.

In order to overcome the above limitation, Zhang, {\it et al} solved the model~(\ref{eq:hamiltonian}) within Hartree-Fock approximation and accounted for various magnetic states~\cite{Zhang2012}. It was shown that it is the distinct band dispersion in both orbitals that can be identified as the crucial ingredient for the presence of orbital selective phase transition with magnetic order.

Since Hartree-Fock approximation usually overestimates the magnetic state, one has to first check the validity of the mean-field calculations by comparing the results with those obtained from the dynamical mean field theory calculations. For this comparison, the chemical potential rather than the filling is fixed as is usually done in the dynamical mean field theory studies, and only the N\'eel antiferromagnetic state is allowed as required by a two-sublattice dynamical mean field theory. Fig.~\ref{fig:DMFT_diffbandstruct_DOSComp}~(e) shows the sublattice magnetization as a function of interaction $U/t$ for the case $t_\alpha=1$, $t_\alpha'=0.6$ and $t_\beta=1$, $t_\beta'=0$. It is found that while the magnetic phase transition obtained from the Hartree-Fock approximation happens earlier than that from the dynamical mean field theory calculations and magnetization is stronger in the Hartree-Fock approximation -indicating that dynamical fluctuations ignored in the Hartree-Fock approximation strongly suppress the magnetically ordered states- the variation of the magnetization as a function of $U/t$ obtained from the dynamical mean field theory calculations can be qualitatively reproduced by the results from the Hartree-Fock approximation. Furthermore, all the phases given from the dynamical mean field theory calculations can be qualitatively captured by the Hartree-Fock approximation as shown in Figs.~\ref{fig:DMFT_diffbandstruct_DOSComp} (f)-(h) and (i)-(k) where the density of states in the different phases are presented. The orbital selective phase is clearly detected by both methods (see Fig.~\ref{fig:DMFT_diffbandstruct_DOSComp}~(g) and Fig.~\ref{fig:DMFT_diffbandstruct_DOSComp}~(j)), implying the validity of the following discussion on the orbital selective phase as well as on other phases in the model~(\ref{eq:hamiltonian}) at the mean-field level. Combining the comparisons at other hopping parameters with different $t_\alpha'$, it was concluded that dynamical fluctuations play a minor role in the orbital selective phase transition.

\begin{figure}[tb]
\includegraphics[width=0.96\textwidth]{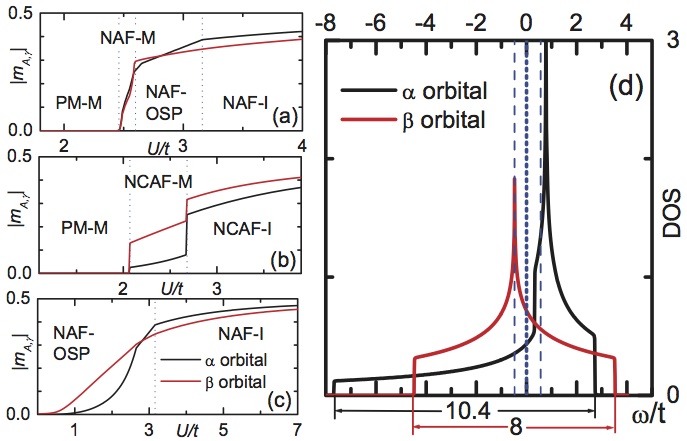}
  \begin{center}
\caption{(Color online) Variation of the magnetization as a function of $U/t$ (a) at $t_\alpha=0.769$, $t_\alpha'=0.615$ and
  $t_\beta=1$, $t_\beta'=0$, (b) at $t_\alpha=1$, $t_\alpha'=0.8$ and $t_\beta=0.769$, $t_\beta'=0.615$, (c) after eliminating the orbital order by adding an effective crystal field splitting of
  $\Delta=0.798$ at $t_\alpha=0.769$, $t_\alpha'=0.615$ and $t_\beta=1$, $t_\beta'=0$, where the term for crystal field splitting is written as $\sum_{i} \Delta (n_{i\beta}-n_{i\alpha})$. Regions of different phases are indicated by the abbreviations, for instance, paramagnetic by PM, N\'{e}el antiferromagnetic by NAF, noncollinear antiferromagnetic by NCAF, while orbital selective phase by OSP. M (I) denotes metal (insulator). Here $J/U=0.25$ and $U'/U=0.5$. (d) Density of states
  at $t_\alpha=1.0$, $t_\alpha'=0.8$, $t_\beta=1$, $t_\beta'=0$, and $U/t=0.0$. Vertical dotted line denotes the Fermi level at half filling, while vertical dashed lines are the "Fermi level" if two orbitals
  are half filled separately. The bandwidths for both two orbitals are also shown here. (From Ref. 7.)} \label{fig:HF_diffbandstruct_DOS}
  \end{center}
\end{figure}

Also in this work, it was found that
 an orbital selective phase transition can still occur at fixed filling of $1/2$ even if various magnetically ordered states as shown in Fig.~\ref{fig:pattern} are present.
 Here, the ground state is obtained after comparisons of the total energies of different magnetic states. Fig.~\ref{fig:HF_diffbandstruct_PD}~(a)-(c) shows the phase transitions happening at $t_\alpha=1$, $t_\alpha'=0.8$ and $t_\beta=1$, $t_\beta'=0$ as a function of $U/t$. As long as $U/t<2.88$, the ground state is a paramagnetic metal with orbital order. In a small interaction region of $2.88<U/t<3.08$, a N\'eel antiferromagnetic metal with orbital order appears. Further increasing $U/t$ from $3.08$ up to $4.02$, the $\alpha$ orbital becomes a N\'eel antiferromagnetic insulator while the $\beta$ orbital remains a N\'eel antiferromagnetic metal, indicating an orbital selective phase. Orbital order disappears in this interaction region. At $U/t>4.02$, both orbitals display N\'eel antiferromagnetic insulating behavior. A phase diagram in the $U/t$-$t_\alpha'$ plane at $t_\alpha=1$, $t_\beta=1$, and $t_\beta'=0$ is presented in Fig.~\ref{fig:HF_diffbandstruct_PD}~(d). An orbital selective phase exists in a wide region of the phase diagram. The phase transitions from both N\'eel antiferromagnetic states to the noncollinear antiferromagnetic state and from paramagnetic metals to N\'eel antiferromagnetic insulators are of first order (solid line), otherwise second order (dotted line).

Though it has been shown that the orbital selective phase transition is still present at fixed filling in finite dimension, the mechanism for it cannot be identified due to the fact that various possible origins coexist. After analyzing the noninteracting electronic density of states (see Fig.~\ref{fig:HF_diffbandstruct_DOS}~(d)), it was found that three factors appear simultaneously in the above study: (I), two orbitals having different bandwidths with the ratio of $W_\alpha/W_\beta=1.3$; (II), the existence of orbital order due to the different band dispersions of the two orbitals ($t_\alpha'/t_\alpha \neq t_\beta'/t_\beta$) which can be viewed as the existence of an effective crystal field splitting; and (III), two orbitals having distinct band dispersions ($2t_{\gamma}(cos k_x +cos k_y) + 4t_{\gamma}' cos k_x cos k_y$) which leads to different shapes of the noninteracting partial DOS. The last effect was not considered in all the previous dynamical mean field studies related to the orbital selective phase transition.

In order to identify the essential mechanism responsible for the orbital selective phase transition observed above, three cases were studied separately:

(I), the effect of different bandwidths was first eliminated by rescaling the hopping parameters of the $\alpha$ orbital from $t_\alpha=1$, $t_\alpha'=0.8$ to $t_\alpha=0.769$, $t_\alpha'=0.615$ so that the ratio of $t_\alpha'/t_\alpha=0.8$ is retained while the ratio of bandwidths becomes $W_\alpha/W_\beta=1$. Fig.~\ref{fig:HF_diffbandstruct_DOS}~(a) presents the various phases as a function of $U/t$ after rescaling. Though the critical points are changed due to the change of the total bandwidths, all the phases involving orbital selective phase are preserved, indicating that such an orbital selective phase transition exists in the absence of bandwidth differences between orbitals.

(II), the orbital order was removed by adding an effective crystal field splitting, by which the half-filling condition is simultaneously satisfied at $U/t=0$ in both orbitals.
Fig.~\ref{fig:HF_diffbandstruct_DOS}~(c) shows that the orbital selective phase transition is still present in the absence of orbital order. However, the states with metallic behavior in both
orbitals vanish since the Fermi level is located right at the van Hove singularity in the $\beta$ orbital at $U/t=0$. It has been checked that a small $t_\beta'$, which shifts the van Hove singularity
away from the Fermi level, leads to the appearance of metallic phases in both orbitals at finite $U/t$.

(III), the effect of orbitals having distinct band dispersions was eliminated but the difference in bandwidth was retained by choosing $t_\alpha=1$, $t_\alpha'=0.8$ and $t_\beta=0.769$,
$t_\beta'=0.615$ which leads to $t_\alpha'/t_\alpha=t_\beta'/t_\beta=0.8$ and $W_\alpha/W_\beta=1.3$. As shown in Fig.~\ref{fig:HF_diffbandstruct_DOS}~(b), an orbital selective phase is precluded by noncollinear antiferromagnetic states, resulting in only two successive phase transitions from paramagnetic metals to noncollinear antiferromagnetic insulators through noncollinear antiferromagnetic metals in both orbitals. Clearly, the orbital selective phase will be replaced by N\'eel antiferromagnetic insulators in both orbitals at any finite $U/t$ if $t_\alpha=1$, $t_\alpha'=0$ and $t_\beta=1.3$, $t_\beta'=0$, which means a similar dispersion relation $t_\alpha'/t_\alpha=t_\beta'/t_\beta=0$ but different bandwidth $W_\alpha/W_\beta=1.3$, since the Fermi level crosses the van Hove singularities in both orbitals.

The results clearly showed that the orbital selective phase transition can occur as long as orbitals have distinct band dispersions even though all the other mechanisms mentioned above are absent while different bandwidth alone will not support the existence of the orbital selective phase transition when magnetic order is considered.

\begin{figure}[tb]
  \begin{center}
\includegraphics[width=0.96\textwidth]{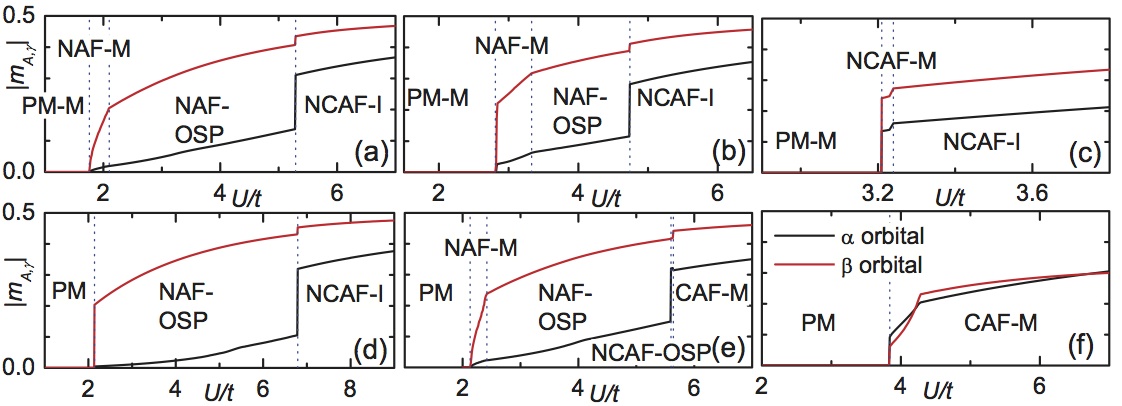}
\caption{(Color online) Variation of the magnetization as a function of $U/t$ at $t_\alpha=1$, $t_\alpha'=2$, $t_\beta=1$. (a) $t_\beta'=0$, $J/U=0.25$, (b) $t_\beta'=0.4$, $J/U=0.25$, (c) $t_\beta'=0.8$, $J/U=0.25$, and (d) $t_\beta'=0$, $J/U=0.0625$ at half-filling. (e) $2.5\%$ and (f) $20\%$ electronic doping at $t_\beta'=0$, $J/U=0.25$. Regions of different phases are indicated by the abbreviations defined, for instance, paramagnetic by PM, N\'{e}el antiferromagnetic by NAF, noncollinear antiferromagnetic by NCAF, collinear antiferromagnetic by CAF, while orbital selective phase by OSP. M (I) denotes metal (insulator). (From Ref. 7.)} \label{fig:HF_diffbandstruct_othereffects}
  \end{center}
\end{figure}

Finally, various effects were investigated on the observed orbital selective phase transition. From the phase diagram  shown in Fig.~\ref{fig:HF_diffbandstruct_PD}~(d), it is obvious that one should discuss two cases separately: 1) $t_\alpha'/t_\alpha>1$ where different magnetic orders like N\'eel and noncollinear antiferromagnetic orders compete with each other; 2) $t_\alpha'/t_\alpha<1$ where only N\'eel antiferromagnetic order occurs.
Fig.~\ref{fig:HF_diffbandstruct_othereffects}, shows the results at $t_\alpha=1$, $t_\beta=1$ and $t_\alpha'=2$. The effect of adding nearest neighbor hopping $t_\beta'$ is first presented. It is found that increasing $t_\beta'$ favors the noncollinear antiferromagnetic state, which squeezes the region of the N\'eel antiferromagnetic orbital selective phase. As seen in Fig.~\ref{fig:HF_diffbandstruct_othereffects}~(b), at $t_\beta'=0.4$, the region of an orbital selective phase is smaller than that at $t_\beta'=0$ (Fig.~\ref{fig:HF_diffbandstruct_othereffects}~(a))
and at $t_\beta'=0.8$ the orbital selective phase completely vanishes (see Fig.~\ref{fig:HF_diffbandstruct_othereffects}~(c)). However, for the case of $t_\alpha'=0.8$, the region of the orbital selective phase remains unchanged at $t_\beta'=0.4$, while it is reasonably replaced by the noncollinear antiferromagnetic state at $t_\beta'=0.8$. The effect of Hund's rule coupling is presented in
Fig.~\ref{fig:HF_diffbandstruct_othereffects}~(d). Compared to Fig.~\ref{fig:HF_diffbandstruct_othereffects}~(a) where $J/U=0.25$, the region of orbital selective phase is enlarged at $J/U=0.0625$ and a direct first-order phase transition from paramagnetic metals in both orbitals to the N\'eel antiferromagnetic orbital selective phase is observed instead of two successive second-order phase transitions through an intermediate N\'eel antiferromagnetic metal at $J/U=0.25$. For the case of $t_\alpha'=0.8$, $t_\beta'=0$, a similar effect of the Hund's rule coupling is found.

Fig.~\ref{fig:HF_diffbandstruct_othereffects}~(e) shows that at a small concentration of electronic doping of $2.5\%$, the orbital selective phase with N\'eel antiferromagnetic order is slightly moved to higher values of $U/t$ and the noncollinear antiferromagnetic insulators existing in the undoped case is replaced by a small region of orbital selective phase with noncollinear antiferromagnetic order which eventually becomes collinear antiferromagnetic metals at larger $U/t$. At large doping of $20\%$, only two phases with paramagnetic state and collinear antiferromagnetic metals remain and the orbital selective phase vanishes as seen in Fig.~\ref{fig:HF_diffbandstruct_othereffects}~(f). The critical value of doping concentration where the orbital selective phase disappears is around $13.6\%$. Similar situation also occurs in the case of $t_\alpha'=0.8$, $t_\beta'=0$. It is interesting to notice that collinear antiferromagnetic metallic states only appear when the system is doped. $20\%$ electronic doping is related to the filling factor in the pnictides where $6$ $3d$ electrons occupy $5$ $3d$ orbitals. However, at this doping rate, the orbital selective phase disappears. Therefore, further investigations on the origin of the possible coexistence of local and itinerant electrons in the iron-based superconductors are required.

%============================================
\subsubsection{Other mechanisms}
\label{sec:three-two-three}
%============================================

\begin{figure}
\begin{center}
\includegraphics[width=0.80\textwidth]{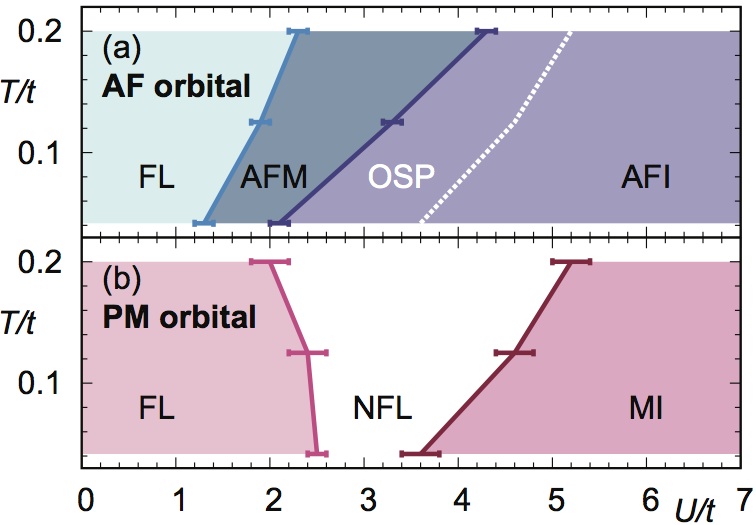}
\end{center}
\caption{(Color online) Left panel: The $U-T$ phase diagram of (a) antiferromagnetic (AF) and (b) paramagnetic (PM) orbitals for $J=U/8$ and $U'=3U/4$.
 The non-Fermi liquid and AF metal (AFM) are determined by the self-energy neither showing divergent behavior nor approaching zero as the Matsubara frequency goes to zero and the spin-dependent symmetry breaking field remaining finite in the metallic state, respectively. Here, FL denotes Fermi liquid, NFL denotes non-Fermi liquid, AFI is antiferromagnetic insulator, MI means Mott insulator, and OSP represents orbital selective phase. (From Ref. 71.)}
 \label{fig:othermechanisms}
\end{figure}

In Sec.~\ref{sec:three-two-two}, it is revealed that unequal band dispersion in different orbitals is one of the mechanisms for the orbital selective phase transition. However, in the previous investigations, the spatial fluctuations were completely ignored. These effects become dominant when the system is highly frustrated, leading to a spin liquid state at large onsite Coulomb interaction $U$. In order to understand the effect of spatial fluctuations when unfrustrated and highly frustrated bands are coupled,  Lee {\it et al.} considered the
 dynamical cluster approximation with 4-site clusters
and the  weak coupling continuous-time quantum Monte Carlo method to
solve a two-orbital Hubbard model without spin-flip term on the
square lattice~\cite{Lee2011}.  To avoid the negative sign problem which is severe in the highly frustrated system, the frustrated band was
 replaced by an unfrustrated band with the constraint that
the  solution should be paramagnetic. This
 can be effectively viewed as a highly frustrated band where magnetic order is completely suppressed by frustration. The second orbital was allowed to have an antiferromagnetic solution. Equal bandwidths are set in both orbitals and the half-filled case is studied.

The phase diagram for $J=U/8$ and $U'=3U/4$ in the $U-T$ plane is shown in the left panel of Fig.~\ref{fig:othermechanisms}. Fermi liquid, non-Fermi liquid and Mott insulating behavior are observed in the paramagnetic orbital. The metallic regions are shrunk due to the enhancement of nonlocal antiferromagnetic correlations as temperature decreases.  The metal-to-insulator transition in the two-orbital Hubbard model (see left panel of Fig.~\ref{fig:othermechanisms}~(b)) is of continuous order which is attributed to the coupling between the paramagnetic orbital with spatial antiferromagnetic correlations and the antiferromagnetic orbital with static antiferromagnetic order. Furthermore, in the weak-coupling region, the antiferromagnetic insulator cannot appear due to thermal and orbital fluctuations between paramagnetic and antiferromagnetic orbitals (see left panel of Fig.~\ref{fig:othermechanisms}~(a)). In fact, in the intermediate region the antiferromagnetic orbital shows an antiferromagnetic metal and as the interaction $U/t$ further increases, the antiferromagnetic insulator state appears. The orbital selective phase where a metallic state in the paramagnetic orbital and an insulating state in the antiferromagnetic orbital coexist, is clearly observed in the intermediate regime. Such a phase is induced by different magnetic states in the two orbitals.

Finally, we will discuss two more
mechanisms proposed
 by Medici {\it et al.}~\cite{Medici2009} and Werner {\it et al.}~\cite{Werner2007}. By applying dynamical mean field theory with the
strong coupling continuous time quantum Monte Carlo method to a two orbital Hubbard model with pair hopping term, Werner and Millis~\cite{Werner2007} reported that away from half filling, an orbital selective phase transition occurs in the presence of a crystal field splitting. When the chemical potential is shifted, the electrons first fill one of the bands and leave the other in a Mott insulating state with a magnetic moment, indicating the presence of an orbital selective phase. Here conditions like the different bandwidths and the integer number of electrons are not required.

Medici {\it et al.}~\cite{Medici2009} proposed another completely different scenario for the orbital selective phase transition. Here the number of orbitals should be larger than 3. For example, in a three-orbital system, with a suitable crystal field splitting, three orbitals split into one orbital which has different energy with respect to the other two, that remain degenerate (or nearly degenerate). In this case, an orbital selective phase transition can take place even in the case of equal bandwidths and Coulomb repulsion for all bands, and for commensurate fillings. This phenomenon can be understood on the basis of  Mott transitions in degenerate bands. It is known that the critical interaction strength $U_c$ for the Mott transition is large if the degeneracy is larger due to their increased kinetic
energy. For example, in the $SU(N)$-orbital Hubbard model $U_c$ scales with $N$ at large $N$, while, for fixed number of bands, a Mott transition occurs at any integer filling and $U_c$ is largest at half-filling and decreases moving away from it. In Ref. 70, an orbital selective phase is found in a large zone of the parameters $U$ and $J$. The phase diagram is obtained from slave spin mean field calculations and the crystal field splitting is adjusted to always have 1 electron in the lifted band, and 1.5 electrons in each of the degenerate ones.

%============================================
\section{Conclusions}
\label{sec:four}
%============================================
In conclusion, we have reviewed various materials which probably show orbital
 selective phases where itinerant and local electrons coexist at the intermediate strength of the local Coulomb interactions. Though, further experiments have to be done to confirm the existence of the orbital selective phase transition, the mechanism for the phase transition has been extensively explored theoretically. Up to now, five distinct mechanisms have been proposed:

(I) {\it different orbitals with unequal bandwidths for integer filling}. In this case, various studies have pointed out that the orbital selective phase transition can occur in spite of whether the spin flip term or pair hopping term is taken into account. However, the nonlocal correlations  induce novel
 momentum selective behavior with new phases. The Hund's rule coupling plays an essential role in the transition which is also affected by various factors like hybridization between orbitals, the crystal field splitting and the ratio of bandwidth~\cite{Ferrero2005,Medici2005,Song2009}.

(II) {\it different orbitals with different band dispersions}. It has been shown that even in the absence of crystal field splittings or when bandwidths, orbital degeneracies, magnetic states and intra-band Coulomb repulsion are equal for different orbitals, the orbital selective phase transition can still occur at different band fillings with magnetic order, as long as different orbitals have distinct band dispersions. This mechanism is in fact very general since usually the strength of hybridization between neighboring sites in different directions is strongly orbital-dependent in real materials, leading to distinct band dispersions in different orbitals. Importantly, the orbital selective phase transition according to the mechanism occurs in a wide range of model parameters, suggesting that this mechanism could be realized in nature.

(III) {\it different orbital with different magnetic states}. The orbital selective phase transition induced by this mechanism is also independent of whether there is the Hund's rule coupling or if the bandwidth is different.

(IV) {\it different orbital degeneracies}. With the help of a suitable crystal field splitting which divides the orbitals into manifolds of different degeneracy and makes at least the orbitals with less degeneracy filled by integer number of electrons, the orbital selective phase transition happens at intermediate local Coulomb interaction if the Hund's coupling is large enough. Here the Hund's rule coupling leads to a band decoupling.

(V) {\it crystal field splitting at non-integer filling}. Changing the filling away from half-filled case can drive the two-orbital system into an orbital selective phase in the presence of a crystal field splitting. This is due to the fact that the band with lower on-site potential will first accommodate additional electrons while that with higher on-site potential remains half-filled.

Since the above mechanisms are quite general and fairly easy to realize, we
conclude that  orbital selective phases should be very common in nature, though it have not been unambiguously detected by experiments.

\section*{Acknowledgments}

Y.Z is supported by National Natural Science Foundation of China (No. 11174219), Shanghai Pujiang Program (No. 11PJ1409900), Research Fund for the Doctoral Program of Higher Education of China (No. 20110072110044) and the Program for Professor of Special Appointment (Eastern Scholar) at Shanghai Institutions of Higher Learning as well as the Scientific Research Foundation for the Returned Overseas Chinese Scholars, State Education Ministry. H.L., H.O.J., and R. V. gratefully acknowledge financial support from the Deutsche Forschungsgemeinschaft through grants FOR 1346 and SFB/TRR 49.

\end{document}